\documentclass[twocolumn]{aastex631}
\usepackage{amsmath}

\defcitealias{Zhou2023}{Z23}
\defcitealias{Benitez-Llambay2023}{BLN23}
\defcitealias{Benitez-Llambay2020}{BLF20}
\defcitealias{Benitez-Llambay2017}{BL17}
\defcitealias{Sawala2016}{Sawala et al. 2016}
\defcitealias{Nebrin2023}{Nebrin et al. 2023}  
\defcitealias{Karunakaran2024}{KS24}

\begin{document}

\title{Not So Round: VLA Observations of the Starless Dark Matter Halo Candidate Cloud-9}

\author[0000-0001-8261-2796]{Alejandro Ben\'itez-Llambay}
\affil{Dipartimento di Fisica G. Occhialini, Universit\`a degli Studi di Milano Bicocca, Piazza della Scienza, 3 I-20126 Milano MI, Italy}

\author[0000-0002-6095-7627]{Rajeshwari Dutta}
\affiliation{IUCAA, Postbag 4, Ganeshkind, Pune 411007, India}

\author[0000-0001-6676-3842]{Michele Fumagalli}
\affil{Dipartimento di Fisica G. Occhialini, Universit\`a degli Studi di Milano Bicocca, Piazza della Scienza, 3 I-20126 Milano MI, Italy}
\affil{INAF – Osservatorio Astronomico di Trieste, via G. B. Tiepolo 11, I-34143 Trieste, Italy}

\author[0000-0003-3862-5076]{Julio F. Navarro}
\affil{Department of Physics and Astronomy, University of Victoria, Victoria, BC V8P 5C2, Canada}

\begin{abstract}
Observations with FAST recently detected H\,{\sc i} 21-cm emission near M94, revealing an intriguing object, Cloud-9, without an optical counterpart. Subsequent analysis suggests Cloud-9 is consistent with a gas-rich ($M_{\rm HI} \approx 10^{6} \ M_{\odot}$), starless dark matter (DM) halo of mass $M_{200} \approx 5 \times 10^{9} \ M_{\odot}$. Using VLA in D-array configuration, we present interferometric observations of Cloud-9 revealing it as a dynamically cold ($W_{50} \approx 12 \rm \ km \ s^{-1}$), non-rotating, and spatially-asymmetric system, exhibiting gas compression on one side and a tail-like structure towards the other, features likely originating from ram pressure. Our observations suggest Cloud-9 is consistent with a starless $\Lambda$CDM dark matter halo if the gas is largely isothermal. If interpreted as a faint dwarf, Cloud-9 is similar to Leo T, a nearby gas-rich galaxy that would fall below current optical detection limits at Cloud-9's distance ($d\approx 5 \rm \ Mpc$). Further observations with HST reaching magnitudes $m_{g} \approx 30$ would help identify such a galaxy or dramatically lower current limits to its stellar mass ($M_{\rm gal} \lesssim 10^{5} \ M_{\odot}$). Cloud-9 thus stands as the firmest starless DM halo candidate to date or the faintest galaxy known at its distance.
\end{abstract}

\keywords{Dark matter(353) --- Cosmology(343) --- Reionization(1383) --- }

\section{Introduction}
\label{Sec:Introduction}
The existence of myriads of starless dark matter halos is a basic tenet of the $\Lambda$ cold dark matter ($\Lambda$CDM) model of structure formation~\citep{Blumenthal1984, Hoeft2006, Okamoto2008, Okamoto2009}. Indeed, reconciling the abundance of collapsed dark matter halos with that of faint galaxies requires stars to form only in a fraction of the halos, particularly those whose mass exceeds a redshift-dependent critical mass, $M_{\rm crit, z}$~\citep[e.g.,][]{Bullock2000, Benson2002}. Within this picture, the only halos that can harbor luminous galaxies in their center today are those whose mass exceeded $M_{\rm crit}$ at some point in the past. On the other hand, halos with masses consistently below this threshold {\it at all times} remain dark~\citep[][and references therein; hereafter BLF20]{Benitez-Llambay2020}. 

Theoretical models and results from numerical simulations suggest a present-day value for the critical mass, $M_{\rm crit,0} \approx 5 \times 10^9 \ M_{\odot}$~\citepalias{Benitez-Llambay2020, Nebrin2023}, indicating that {\it all} dark matter halos with virial mass, $M_{200} > M_{\rm crit,0}$, should host luminous galaxies today. On the other hand, halos with masses below this value may or may not host stars, depending on their particular mass assembly history and on whether they exceeded $M_{\rm crit}$ in the past.

Using high-resolution hydrodynamical cosmological simulations,~\citet[][hereafter BL17]{Benitez-Llambay2017} studied the origin and detailed properties of starless halos. They found that the most massive systems, with masses comparable to $M_{\rm crit}$ at present, still retain gas confined to their center. Despite significant gas loss due to the impact of the external ultraviolet background (UVB) radiation field responsible for reionizing the universe, these halos maintain a small gas reservoir in or near hydrostatic equilibrium with the dark matter halo and in thermal equilibrium with the UVB. The relatively high density and low temperature of the gas in these systems facilitate the recombination of hydrogen, resulting in the formation of a prominent neutral hydrogen (H\,{\sc i}) core in the center. This is why we call these massive H\,{\sc i}-rich starless halos Reionization-Limited-HI-Clouds (RELHICs).

RELHICs should display distinctive properties: they should contain a central nearly-spherical H\,{\sc i} core in hydrostatic equilibrium and thermal equilibrium with the UVB, making them detectable through their narrow 21-cm emission line. These properties make RELHICs remarkable tests of the $\Lambda$CDM model on the smallest scales because their detailed gaseous structure offers a direct probe into the underlying structure of their pristine dark matter halos.

Recent observations by~\citet[][hereafter, Z23]{Zhou2023} with the Five-hundred-metre Aperture Spherical Telescope (FAST) detected marginally resolved 21-cm emission near the M94 galaxy, separated by $\approx 50'$ in projection, without an optical counterpart. Unlike previous candidate ``dark'' H\,{\sc i} clouds~\citep[e.g.][]{Adams2013, Xu2023}, all known properties of this source (Cloud-9 hereafter) appear broadly consistent with RELHICs, as pointed out by~\citetalias{Zhou2023} and recently discussed by~\citet[][hereafter, BLN23]{Benitez-Llambay2023}. Subsequent unresolved observations with the Green Bank Telescope (GBT) confirmed the detection of this radio source~\citep[][hereafter KS24]{Karunakaran2024} but with a slightly broader and more asymmetric line profile than that reported by~\citetalias{Zhou2023}. Although the observations of~\citetalias{Karunakaran2024} suggest Cloud-9 has a more complex structure than anticipated, they could not exclude Cloud-9 from being a RELHIC. Assuming a distance to the object, $d = 4.66 \rm \ Mpc$\footnote{We refer the reader to the work of~\citetalias{Benitez-Llambay2023} for a detailed discussion on the distance estimate for Cloud-9.}, FAST and GBT observations bracket the total H\,{\sc i} mass of the system in the range, $1 \times 10^{6} \lesssim M_{\rm HI} / M_{\odot} 
\lesssim 2 \times 10^6$. Moreover, at the adopted distance, Cloud-9 must be dark matter dominated, since its inferred dynamical mass is much higher than its  H\,{\sc i} mass~\citepalias{Benitez-Llambay2023}.

\begin{figure*}
        \centering
	\includegraphics{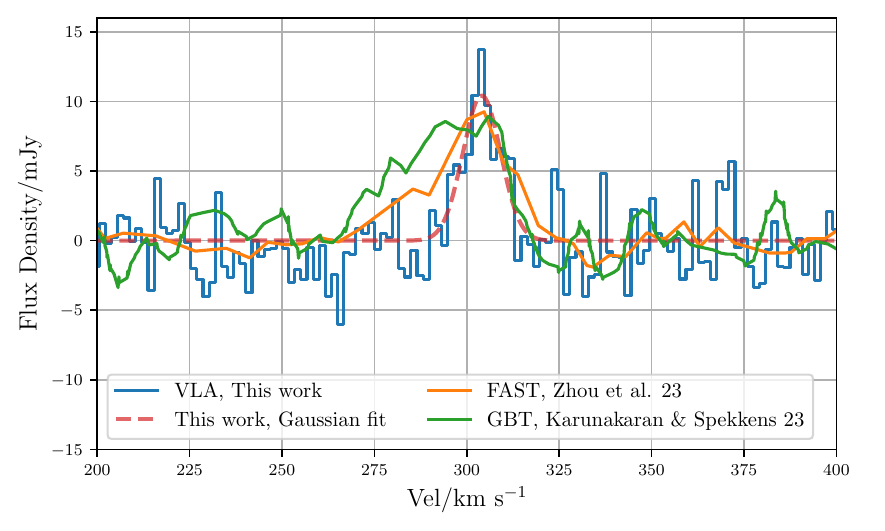}
        \caption{VLA-D spectrum of Cloud-9 centered at the H\,{\sc i} 21-cm line (blue solid histogram), compared with the spectra obtained by FAST (orange solid line) and GBT (green solid line). The red dashed line shows a Gaussian fit to the VLA-D line profile centered at $v_{0} \approx 304 \rm \ km \ s^{-1}$, and full width at half maximum, $W_{\rm 50} \approx 12 \ \rm km \ s^{-1}$. Both FAST and GBT data were taken from the work of~\citetalias{Karunakaran2024}.}
    \label{Fig:Figure1}
\end{figure*}

Although Cloud-9 is overall consistent with a RELHIC, the main doubts about its true nature concern primarily the low angular resolution of the FAST observations (half-power beamwidth $\approx 2.9 \rm \ arcmin$ at 1.4 GHz). FAST observations thus offer limited insight into the system's central H\,{\sc i} column density, making it difficult to impose tight constraints on its size and projected shape. In addition, Cloud-9's detailed column density profile appears slightly, but systematically, more extended than expected for RELHICs. This mismatch could originate for several reasons ranging from uncertainties in the FAST primary beam's characterization to the system being closer to us, to even departures from the $\Lambda$CDM model~\citepalias{Benitez-Llambay2023}. Characterizing the system better thus requires higher-resolution radio observations.

On the other hand, using the surface brightness limits from the DESI Legacy Imaging Survey (DESI LS) estimated by~\cite{Martinez-Delgado2023} –$29.15$, $28.73$, and $27.52$ mag arcsec$^{-2}$ in the $g$, $r$, and $z$ bands, respectively–,~\citetalias{Zhou2023} derived a large upper limit to the stellar mass of any luminous counterpart (galaxy mass, $M_{\rm gal} \lesssim 10^{5} \ M_{\odot}$). This particularly high galaxy mass opens the possibility that Cloud-9 is an analog of some of the faint Local Group dwarf galaxies but at the M94 distance. Elucidating whether or not Cloud-9 has a luminous counterpart thus requires deeper observations in the optical.

In this paper, we aim to mitigate current limitations by considering interferometric observations of Cloud-9 in radio wavelengths. To this end, we use the Karl G. Jansky Very Large Array (VLA) in the D-array configuration (hereafter VLA-D). This configuration provides higher spatial resolution than FAST and GBT and allows resolving the central parts of the system better in comparison. We start by discussing the observations and data reduction in Section~\ref{Sec:Observations}. We then present the results and discussions in Section~\ref{Sec:Results} and Section~\ref{Sec:Discussion}, respectively, and summarize our main results and conclusions in Section~\ref{Sec:Conclusions}.
%__________________________________________________________________

\section{Observations and data reduction}
\label{Sec:Observations}

\subsection{Data acquisition}

We observed Cloud-9 with the L-band receivers of VLA in the D-array configuration between 1 and 13 December 2023 (Proposal ID: VLA/23B-313; PI: A. Benitez-Llambay). The observations were centered at RA, Dec (J2000) = 12h 51m 52s, +40d 17m 29s. The total on-source integration time was $\approx 3.2$ h, which was split into four separate observing runs. The primary calibrator, 3C286, was observed once in each observing run for calibration of flux density and bandpass. In addition, a secondary calibrator, J1227+3635, was observed at regular intervals during the observing runs for complex gain calibration. The observations used the WIDAR correlator with the 8-bit samplers, two polarizations, and a single 8 MHz spectral window that was centered on the redshifted H\,{\sc i} 21-cm line. This configuration yielded a velocity coverage of $\approx$1700 km\,s$^{-1}$ at Cloud-9's redshift, adequate to cover the H\,{\sc i} 21-cm emission detected by FAST, and to provide line-free channels for continuum subtraction. The spectral window was split into 1024 channels, resulting in a spectral resolution of 7.8 kHz per channel or $\approx 1.7$ km\,s$^{-1}$, similar to that of the FAST spectrum reported by~\citetalias{Zhou2023}.

\subsection{Data reduction}

The data were reduced using the Common Astronomy Software Applications package \citep[{\sc casa}, version 6.6;][]{McMullin2007}, following standard procedures. The dataset of each observing run was calibrated independently for the flux density scale, bandpass, and complex gain, after flagging the bad data due to non-working antennae and radio frequency interference. The continuum was subtracted in the $uv$-plane using a first-order polynomial fit to the line-free channels (excluding edge channels). The continuum-subtracted visibilities of each observing run were combined to produce the H\,{\sc i} image cube with the routine {\tt TCLEAN} in {\tt CASA}, using a cell size of 9 arcsec and Natural weighting. The cube was ``cleaned'' during the imaging process using auto-masking by multiple thresholds for the deconvolution. Any residual continuum was subtracted from the image cube using a first-order polynomial fit to the line-free channels (excluding edge channels). The resulting synthesized beam of the final cube has a size $65.2''\times53.7''$ with a position angle of $\approx -76^{\rm o}$, and an $rms$ uncertainty of $\approx 1.3$ mJy\,beam$^{-1}$ per velocity channel of 1.7 km\,s$^{-1}$.
%__________________________________________________________________
\section{Results}
\label{Sec:Results}

\subsection{Cloud-9's VLA-D Spectrum}

\begin{figure}
	\includegraphics[width=\columnwidth]{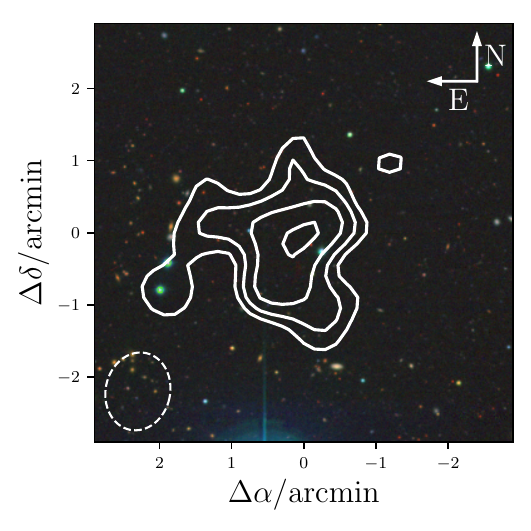}
        \caption{Cloud-9's H\,{\sc i} column density isocontours superimposed on an image cutout of the DESI Imaging Legacy Survey. The ellipse at the bottom left indicates the ``clean''  beam. The isocontours start at $3\sigma_{\rm N_{\rm HI}} = 5.45 \times 10^{18} \ \rm cm^{-2}$, where $\sigma_{\rm N_{\rm HI}}$ is the {\it rms} uncertainty, and they increase in steps of $\sqrt{2}$ towards the centre. The maximum column density reached in the centre is $N_{\rm HI,0} = 1.6 \times 10^{19} \rm \ cm^{-2}$. The image is centred at $(\alpha, \delta)_{\rm J2000}=(192.9642^{o}, 40.30139^{o})$.}
    \label{Fig:Figure2}
\end{figure}

We show Cloud-9's H\,{\sc i} 21-cm spectrum, specifically the flux density as a function of the velocity channel, in Figure~\ref{Fig:Figure1} (blue line), extracted by integrating, for each channel, the flux contained within the outermost contour shown in Figure~\ref{Fig:Figure2}. For comparison, we also include the spectra obtained from single-dish radio observations carried out with FAST (orange line) and GBT (green line), both taken from Figure 1 of~\citetalias{Karunakaran2024}.

Our VLA-D observations confirm the detection of emission from H\,{\sc i} 21-cm at heliocentric velocities of approximately $v_{\rm helio} \approx 300 \rm \ km \ s^{-1}$, consistent with the results of~\citetalias{Zhou2023} and~\citetalias{Karunakaran2024}. However, unlike the GBT results, our measured line profile appears fairly symmetric and is well described by a Gaussian function. The red dashed line shows the best-fit Gaussian model to the VLA-D line profile and is consistent with a mean heliocentric recessional velocity of $v_{\rm helio, VLA} \approx (304 \pm 2) \rm \ km \ s^{-1}$ and a full width at half maximum of $W_{\rm 50, VLA} \approx (12 \pm 1) \rm \ km \ s^{-1}$. Although the recessional velocity agrees with results from FAST and GBT, our estimate of the line broadening is roughly a factor of two smaller than that obtained by these instruments.\footnote{We note that~\citetalias{Zhou2023} report a line broadening of approximately $\approx 20 \rm \ km \ s^{-1}$ for Cloud-9, but a subsequent data release as part of the FASHI catalogue~\citep{Zhang2024} indicates that Cloud-9's line broadening, as observed by FAST, is consistent with GBT results. This issue was pointed out by~\citetalias{Karunakaran2024}.}

\subsection{Cloud-9's H\,{\sc i} Column Density Map}
To describe the spatial distribution of Cloud-9, we estimate the H\,{\sc i} column density map, assuming, as usual, that the source is optically thin. Under this assumption, the relation between the H\,{\sc i} column density, $N_{\rm HI}$, and the measured surface brightness temperature, $T_{\rm B}$, becomes \citep[e.g.,][]{Dickey1990,McClure-Griffiths2023}:
\begin{equation}
    N_{\rm HI} = (1.823 \times 10^{18} \rm \ cm^{-2}) \displaystyle\int{\left ( \frac{T_{B}}{K} \right )  \left ( \displaystyle\frac{dv}{\rm km \ s^{-1}} \right ) },
\end{equation}
in which the integration is carried out over the velocity channels, and $T_{B}$ is the brightness temperature, related to the observed flux density, $I_{\nu}$, at frequency $\nu$, through \citep[e.g.,][]{Wilson2009}:
 
\begin{multline}
        T_{B} = \left ( 1.68 \times 10^2 \rm \ mK  \right ) \left ( \displaystyle\frac{I_{\nu}}{\rm Jy \ beam^{-1}} \right ) \\ 
        \left ( \displaystyle\frac{\nu}{1.42 \ \rm GHz} \right)^{-2} \left ( \displaystyle\frac{\theta_m}{60''} \right )^{-1}  \left ( \displaystyle\frac{\theta_M}{60''} \right )^{-1}.
\end{multline}

The angles $\theta_m$ and $\theta_M$ correspond to the full width at half maximum of the ``clean'' beam along the minor and major axes, respectively. For reference, with the adopted VLA-D configuration and data reduction pipeline (see Sec.~\ref{Sec:Observations}), the mean ``clean'' beam, averaged over the velocity channels, is well described by $(\theta_m, \theta_M) \approx (53.7'', 65.2'') $, and a position angle, $\rm \beta \approx -76^{o}$ (as shown by the dashed ellipse in the bottom left of Figure~\ref{Fig:Figure2}).

We select the velocity integration range to maximize the signal-to-noise ratio ($\rm S/N$) while ensuring convergence of the maximum column density value across the map. In practice, these criteria lead to an optimal integration range in the interval $(293-314) \rm \ km \ s^{-1}$ (12 channels), resulting in a column density {\it rms} uncertainty, $\sigma_{\rm N_{HI}} \approx 1.8 \times 10^{18} \rm \  cm^{-2}$, and a peak signal-to-noise ratio of approximately $\rm (S/N) \approx 9$. We employ a similar procedure to calculate the source's total flux.

We show the resulting column density map in Figure~\ref{Fig:Figure2}. The white solid lines display four H\,{\sc i} column density isocontours, starting at $3\sigma_{\rm N_{HI}}$ in the outer parts and increasing in steps of $\sqrt{2}$ towards the center. The origin of the map corresponds to the location where the system reaches the peak column density, $N_{\rm HI,0} \approx 1.63 \times 10^{19} \rm \ cm^{-2}$, with coordinates, $(\alpha, \delta)_{\rm J_{2000}} = (12^{\rm h} 51^{\rm m}51.3^{\rm s}, +40^{o}18'5'')$. 

Although the central parts of the system are arguably unresolved given the extent of the ``clean'' VLA-D beam, the higher spatial resolution of our observations reveals a more complex structure than anticipated from the lower resolution FAST dataset (we refer the reader to Figure 1 of~\citetalias{Benitez-Llambay2023} for comparison with the FAST H\,{\sc i} column density map, which looks fairly round and symmetric). 

Unlike the single-dish FAST observations, our interferometric VLA-D data reveal Cloud-9 as a system with a slightly distorted morphology relative to its peak emission. The system is extended towards the east, compact towards the west, and does not entirely resemble a sphere in projection. Interestingly, hints of this asymmetry, particularly the extended emission towards the far east side, were already present in the FAST data.

As we will discuss later in Section~\ref{Sec:Discussion}, the perturbed morphology of Cloud-9 could originate from interactions with the environment, particularly ram pressure, which is expected to compress the gas in one direction and leave behind a tail-like structure, features consistent with the observed gas distribution.

\subsection{Cloud-9's Total H\,{\sc i} Flux}

The total VLA-D flux density of the source within the $3\sigma_{\rm N_{HI}}$ region is $S_{21} \approx (0.13 \pm 0.02) \ \rm mJy \ km \ s^{-1}$. We have verified that this flux value is robust, as measurements over a smoothed image yield comparable results. Specifically, when applying a natural weighting scheme with a Gaussian taper, producing a larger synthesized beam of size $(\theta_{m,t}, \theta_{M,t}) \approx (105'', 86'')$ at PA = -80 degrees, results in a lower flux value of $S_{21,t} \approx (0.1 \pm 0.1) \ \rm mJy \ km \ s^{-1}$.

Our measurement of the flux density value is thus systematically lower than those measured by~\citetalias{Zhou2023} using FAST, $(0.24 \pm 0.01) \rm \ mJy \ km \ s^{-1}$, and~\citetalias{Karunakaran2024} using GBT, $(0.28 \pm 0.04) \rm \ mJy \ km \ s^{-1}$. At Cloud-9's fiducial distance, $d = 4.66 \rm \ Mpc$, the VLA-D integrated flux density corresponds to a total H\,{\sc i} mass, $M_{\rm HI} = \left (7 \pm 1 \right) \times 10^{5} \ M_{\odot}$, which we obtain from the usual relation~\cite[e.g.,][]{Roberts1962}:
\begin{multline}
    M_{\rm HI} = 2.36 \times 10^{5} \ M_{\odot} \left ( \displaystyle\frac{d}{1 \rm \ Mpc}\right )^2 \\ \displaystyle\int{\left ( \displaystyle\frac{I_{\nu}}{\rm mJy} \right) \ \left (\displaystyle\frac{dv}{\rm km \ s^{-1}} \right ).} 
\end{multline}
The overall VLA-D flux (H\,{\sc i} mass) deficit compared with FAST and GBT, together with the discrepancy in the shape of the line profile, may originate from either the presence of diffuse emission missed in our VLA-D interferometric observations or contamination of the GBT field by emission we do not sample; note that the largest angular scale limit imposed by the L-band VLA-D configuration is $\approx 16$ arcmin. Additionally, the relatively lower surface brightness sensitivity and incompleteness in the $uv$-plane coverage, particularly at short baselines, may preclude adequate sampling of large-scale diffuse emissions on scales smaller than this limit. The systematic difference among the datasets obtained with GBT, FAST, and VLA-D suggests that all these issues may be contributing factors~\citep[see, e.g.,][for a recent comparison between FAST and VLA observations showing similar discrepancies]{Wang2024}. Nevertheless, our observations compellingly demonstrate the presence of a prominent neutral phase with a fairly narrow velocity profile. 

\begin{figure}
        \centering
	\includegraphics[width=\columnwidth]{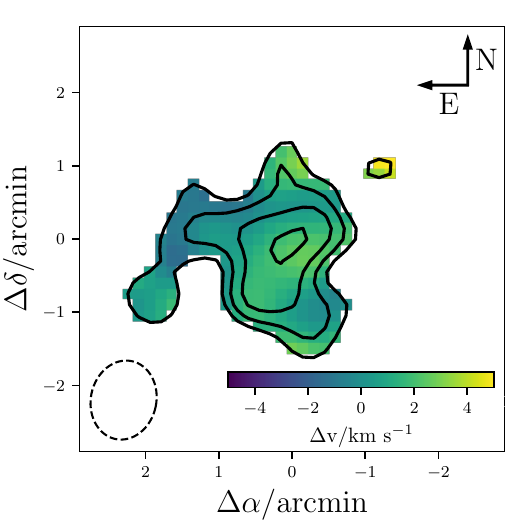}
        \caption{Cloud-9's intensity-weighted velocity map, relative to the inferred heliocentric velocity of the line, $v_{\rm helio, VLA} = 304 \rm \ km \ s^{-1}$. The dashed ellipse in the bottom left indicates the extent of the beam, while the black lines show the same column density isocontours in Figure~\ref{Fig:Figure2}. The lack of a velocity gradient suggests the absence of ordered rotation.}
    \label{Fig:Figure3}
\end{figure}

\subsection{Cloud-9's H\,{\sc i} Intensity-weighted Velocity Map}
To quantify the system's degree of ordered rotation, we consider the intensity-weighted velocity map of the spectral line in Figure~\ref{Fig:Figure3}, in which we only display the regions with column density above $3\sigma_{\rm N_{HI}}$. The color map indicates the velocity relative to the overall inferred heliocentric velocity, $v_{\rm helio,VLA} = 304 \rm \ km \ s^{-1}$. The lack of coherent rotation over the sampled scales is readily apparent, demonstrating that the sampled gas cloud is not supported by circular motion. This is consistent with FAST results, which also reveal Cloud-9 as a non-rotating system.

We thus conclude that the broad properties of Cloud-9, as derived from our VLA-D observations, are still consistent with those expected for a RELHIC, except for its shape. We will discuss the potential origin of this distorted morphology in Section~\ref{Sec:Discussion}. 

\begin{figure}
	\includegraphics[width=\columnwidth]{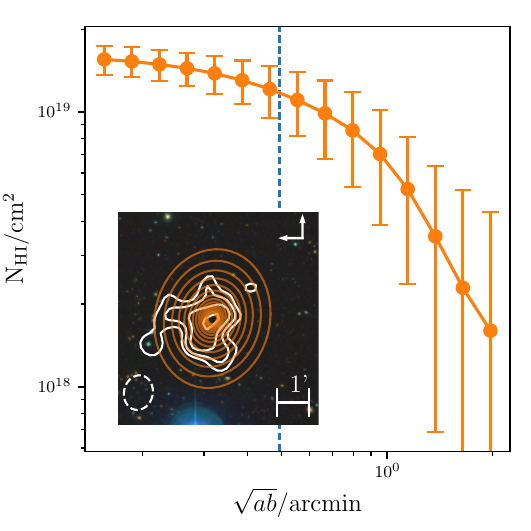}
        \caption{Cloud-9's column density profile as a function of the effective radius, $R_{\rm eff}=\sqrt{ab}$, in which $a$ and $b$ are the minor and major semiaxes of each ellipse shown in the inset, and are in identical proportion to those of the beam. The reported column density corresponds to the mean column density of the pixels intersected by each ellipse. The uncertainties represent the $rms$ scatter along the ellipse and coincide with the typical dispersion found by measuring the profile along independent directions. The vertical dashed line indicates the radial extent of the beam.}
    \label{Fig:Figure4}
\end{figure}

\subsection{Cloud-9's Detailed H\,{\sc i} Column Density Profile}

We now turn our attention to Cloud-9's detailed column density profile. 
Given the asymmetry of the H\,{\sc i} isocontours and the high ellipticity of the VLA-D beam, we measure Cloud-9's detailed column density profile within ellipses centered at the peak. For each ellipse, we calculate the effective radius as the geometric mean of the minor and major semiaxes, $R_{\rm eff} = \sqrt{ab}$, which are in proportion to those of the beam (see the inset panel in Figure~\ref{Fig:Figure4}). We then estimate the column density at $R_{\rm eff}$ as the mean column density of the pixels intersected by each ellipse and the scatter as the standard deviation. We have verified that the scatter obtained in this manner effectively captures the dispersion obtained by measuring the column density profile along various independent directions. For example, measuring the profile from the center towards the west produces a profile consistent with the upper bound. Similarly, measurements towards the east produce a profile consistent with the lower bound. The resulting column density profile together with the associated {\it rms} is shown in Figure~\ref{Fig:Figure4}, in which the vertical dashed line indicates the radial extent of the beam.

\section{The Nature of Cloud-9}

Having presented Cloud-9's properties as observed by VLA-D, we now analyze two possible interpretations for the system. One interpretation is that Cloud-9 is an H\,{\sc i}-rich starless dark matter halo (RELHIC), as predicted by state-of-the-art numerical hydrodynamical simulations, an idea explored qualitatively by~\citetalias{Zhou2023} and quantitatively by~\citetalias{Benitez-Llambay2023}. The other interpretation is that Cloud-9 is an H\,{\sc i}-rich dwarf galaxy that is too faint to be detected by current optical observations of this field. These possibilities are justified by the fact that, as discussed next, a large amount of mass other than neutral hydrogen is required for the system to be largely in equilibrium.

If Cloud-9 is in or near hydrostatic equilibrium in the central regions, the free-fall time should be comparable to the sound crossing time within the observed range. Defining the free-fall time as, $t_{\rm ff} = \left ( 3\pi / 16 G \bar \rho_{\rm HI} \right )^{1/2}$, and the sound-crossing time as $t_{\rm sc} = \left ( r / c_{s} \right )$, with $G$ being Newton's gravitational constant, $\bar \rho_{\rm HI} \sim 3 M_{\rm HI} / 4 \pi r^3$ the mean enclosed neutral hydrogen density within a sphere of radius $r$, and $c_{s}$ the sound speed of the gas, the identity $t_{\rm ff} \sim t_{\rm sc}$ yields:
\begin{equation}
    r \sim \left ( 8.5 \ {\rm pc} \right ) \left ( \displaystyle\frac{M_{\rm HI}}{7 \times 10^{5} M_{\odot}} \right ) \left ( \displaystyle\frac{c_{s}}{12 \ {\rm km \ s^{-1}}} \right )^{-2}.
\end{equation}
At Cloud-9's distance, $8.5 \rm \ pc$ corresponds to a projected size of approximately $0.4 \ \rm arcsec$, which is orders of magnitude smaller than observed. This mismatch implies the necessity for a substantial amount of gravitational mass other than neutral hydrogen to sustain equilibrium over the observed scales. 

\subsection{Cloud-9 as a Starless $\Lambda$CDM RELHIC Candidate}
\label{Sec:RELHIC_models}

To test the RELHIC hypothesis for Cloud-9 we compare its detailed column density profile with that of RELHICs. For this comparison, we use analytic RELHIC models constructed as described by~\citetalias{Benitez-Llambay2023}, a procedure we explain below.

For a given Navarro-Frenk-White~\citep[][hereafter NFW]{Navarro1996} dark matter halo of virial mass, $M_{200}$, and concentration, $c_{\rm NFW}$, we calculate its detailed gas density profile, assuming the gas is in hydrostatic equilibrium with the dark matter halo and in thermal equilibrium with the external UVB, whose spectrum corresponds to that of \cite{Haardt2012}. To integrate the hydrostatic equilibrium equation we assume a boundary condition such that the system's pressure at infinity matches that of the intergalactic medium at the mean density of the universe. This approach, which has no free parameters other than the present-day mean baryon density, the mass of the halo, and the UVB spectrum, has been shown to reproduce results from hydrodynamical simulations \citepalias[e.g.,][]{Benitez-Llambay2017,Benitez-Llambay2020}. We shall refer to this model as the~\citetalias{Benitez-Llambay2017} RELHIC model.

\begin{figure}
\includegraphics[width=\columnwidth]{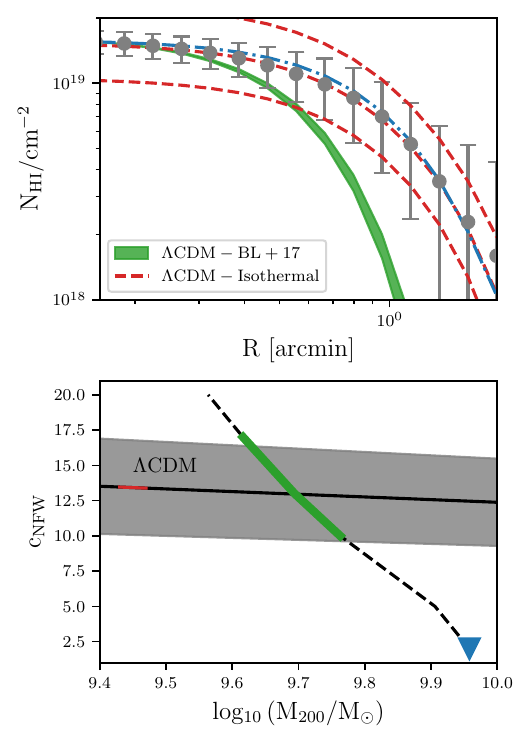}
        \caption{The {\it top panel} shows Cloud-9's H\,{\sc i} column density profile (grey points with error bars), compared with the mock-observed profile of $\Lambda$CDM RELHICs modeled following the~\citetalias{Benitez-Llambay2017} model whose central column density matches Cloud-9's (green shaded region), and isothermal RELHICs with sound speed matching Cloud-9's observed 21-cm line broadening (red dashed lines). These RELHICs inhabit dark matter halos with virial masses and concentrations consistent with the $\Lambda$CDM mass-concentration relation. The blue dot-dashed line represents a~\citetalias{Benitez-Llambay2017} RELHIC in a dark matter halo with $M_{200} \approx 9 \times 10^{10} \ M_{\odot}$ and concentration $c_{\rm NFW}=2$, which is incompatible with $\Lambda$CDM expectations. The {\it bottom panel} shows the $\Lambda$CDM mass-concentration relation with $1\sigma$ scatter (black line with shaded region), along with the combinations of halo mass and concentration that yield RELHIC column density profiles matching Cloud-9's central column density (black dashed line). The thick green/red lines indicate the mass and concentration range producing the green/red profiles in the top panel, while the blue triangle marks the location of the RELHIC shown by the blue line in the top panel.}
    \label{Fig:Figure5}
\end{figure}

To convert the gas density profile into an H\,{\sc i} density profile, we use the results from \cite{Rahmati2013}, which provide the H\,{\sc i} fraction as a function of total hydrogen gas density and temperature. We then integrate the resulting H\,{\sc i} density profile along the line of sight to obtain the intrinsic H\,{\sc i} column density profile. We place this intrinsic profile at Cloud-9's distance and convolve it with the VLA-D ``clean'' beam, resulting in a mock  H\,{\sc i} column density profile that we ``observe'' and measure along the same ellipses defined before.

We note that Cloud-9's distorted morphology, particularly in the outer regions, may indicate that the system is being perturbed. For example, if the system were moving within a gaseous background associated with M94 and experiencing ram pressure, it would display similar features as observed and would likely tend to expand due to the removal of gas in the outer parts. However, whether this gas becomes photoionized by the external photoionizing background as it expands depends on the system's intrinsic gas density. In addition, there is no guarantee that the local intensity of the UVB at Cloud-9's location is that of~\cite{Haardt2012}. 

As shown by~\citetalias{Benitez-Llambay2017} and~\cite{Sykes2019}, the neutral phase of RELHICs can reach temperatures $\lesssim 10^4 \ \text{K}$, which roughly corresponds to a sound speed of about $c_s \approx 9 \ \rm km \ s^{-1}$, assuming a mean molecular weight of $\mu = 1.2$. The broadening of Cloud-9's H\,{\sc i} 21-cm line, as inferred from our observations, is approximately $12 \rm \ km \ s^{-1}$, implying a temperature for Cloud-9 exceeding by more than a factor of 2 the temperature expected for RELHICs. Additionally, the higher H\,{\sc i} 21-cm line broadening measured by FAST~\citepalias{Zhou2023} and GTB~\citepalias{Karunakaran2024}, together with Cloud-9's asymmetric morphology, may indicate the presence of a perturbing mechanism that could also be responsible for thermalizing the system to a higher temperature than expected for RELHICs. 

Because of these reasons, we also explore an alternative ``isothermal'' model, assuming the gas is fully isothermal with sound speed $c_{s} = 12 \rm \ km \ s^{-1}$, a value matching Cloud-9's observed H\,{\sc i} 21-cm line broadening. For reference, this sound speed corresponds to a gas temperature of about $\approx 2 \times 10^{4} \rm \ K$, assuming again a mean molecular weight of $\mu = 1.2$. We shall refer to this model as the ``isothermal'' RELHIC model.

Similar to Figure~\ref{Fig:Figure4}, the top panel of Figure~\ref{Fig:Figure5} shows Cloud-9's detailed column density profile and its associated scatter (grey circles). The green shaded region represents the range of profiles allowed for the~\citetalias{Benitez-Llambay2017} $\Lambda$CDM RELHICs, matching Cloud-9's central maximum column density when placed at Cloud-9's distance and convolved with the VLA-D beam. The bottom panel's green line indicates that these RELHICs reside in halos with virial masses in the range $4 \lesssim M_{200} / 10^{9} \ M_{\odot} \lesssim 6$, and concentrations in the range  $17 \gtrsim c_{\rm NFW} \gtrsim 10$, all consistent with the $\Lambda$CDM mass-concentration relation~\citep[e.g.,][grey shaded region in the bottom panel]{Ludlow2016}. 

The comparison carried out in Figure~\ref{Fig:Figure5} reveals that Cloud-9's column density profile is overall inconsistent with the~\citetalias{Benitez-Llambay2017} RELHIC models. In other words, Cloud-9 cannot be described as a gas cloud in hydrostatic equilibrium with $\Lambda$CDM halos in an otherwise unperturbed background illuminated by the~\cite{Haardt2012} UVB. Specifically, Cloud-9's H\,{\sc i} distribution is too extended to align with this model fully. This inconsistency may partially stem from Cloud-9's distorted morphology, which suggests the system is not truly in isolation. Alternatively, the mismatch may be ascribed to an intrinsic difference in the density or temperature structure of Cloud-9 compared to the~\citetalias{Benitez-Llambay2017} models, or even departures from $\Lambda$CDM, as pointed out by~\citetalias{Benitez-Llambay2023}. We discuss these issues further below and in Sec.~\ref{Sec:Discussion}.

\begin{figure*}
\includegraphics[]{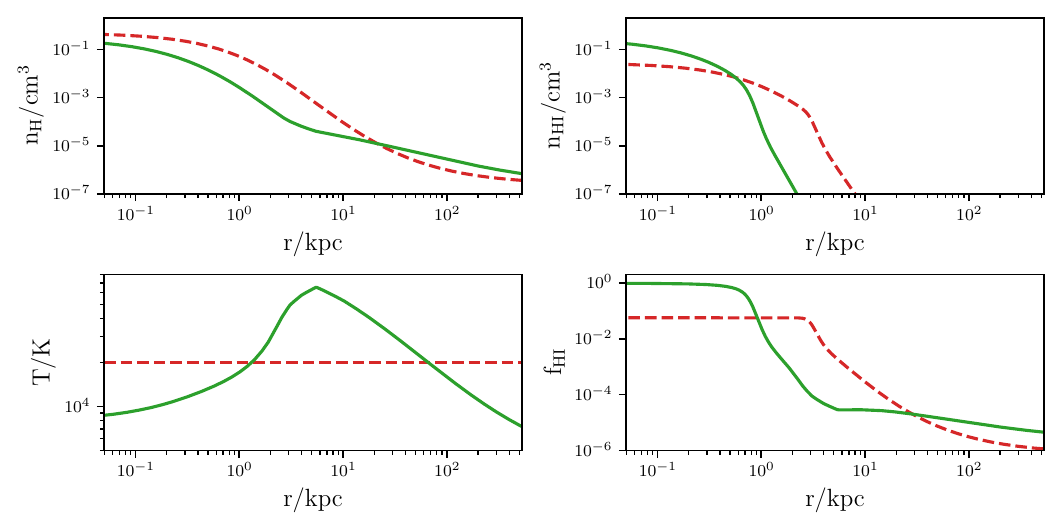}
        \caption{Total (top left) and neutral hydrogen density (top right), gas temperature (bottom left), and H\,{\sc i} fraction (bottom right), as a function of radius. Different lines correspond to the RELHICs shown in Figure~\ref{Fig:Figure5}, specifically: a \citetalias{Benitez-Llambay2017} RELHIC inhabiting a $\Lambda$CDM halo of mass, $M_{200} \approx 5 \times 10^{9} \ M_{\odot}$, matching Cloud-9's central column density (green solid line), and an isothermal RELHIC consistent with observations (red solid line), with mass, $M_{200} \approx 4 \times 10^{9} \ M_{\odot}$, both with concentration, $c_{\rm NFW} \approx 13$.}
    \label{Fig:Figure6}
\end{figure*}

Following~\citetalias{Benitez-Llambay2023}, we now take the discrepancy between Coud-9's column density profile and the~\citetalias{Benitez-Llambay2017} RELHIC model at face value and explore whether it is possible to alleviate the issue by considering a dark matter halo that departs from $\Lambda$CDM expectations. In agreement with~\citetalias{Benitez-Llambay2023}, we find that the discrepancy between the observed column density profile and RELHIC models is resolved if we allow a RELHIC to occupy a dark matter halo with extremely low concentration, so low that it becomes incompatible with $\Lambda$CDM. For instance, the RELHIC represented by the blue dot-dashed line in the top panel of Figure~\ref{Fig:Figure5}, residing in a dark matter halo of mass, $M_{200} \approx 9 \times 10^{9} \ M_{\odot}$, and concentration, $c_{\rm NFW} \approx 2$, aligns with observations. However, this RELHIC inhabits a dark matter halo that significantly deviates from $\Lambda$CDM expectations, as indicated by the blue triangle in the bottom panel, which lies by more than $3 \sigma$ away from the~\cite{Ludlow2016} $\Lambda$CDM mass-concentration relation.

Interestingly, by adopting a gas temperature of $2\times 10^{4} \ K$ ($c_{s} \approx 12 \rm \ km \ s^{-1}$), it is possible to find a family of ``isothermal" RELHICs inhabiting $\Lambda$CDM halos that match Cloud-9's H\,{\sc i} column density profile. Indeed, the red dashed lines in the top panel of Figure~\ref{Fig:Figure5} show three isothermal models with masses in the range $3.9 \lesssim M_{200} / 10^{9} \ M_{\odot} \lesssim 4.3$, and concentration matching the $\Lambda$CDM mass-concentration relation (see the red line in the bottom panel). Remarkably, these RELHICs not only bracket Cloud-9's column density profile but also have a similar shape. Moreover, it is possible to find one model that reproduces observations.  

To understand where the differences between the~\citetalias{Benitez-Llambay2017} and isothermal RELHICs originate, we consider, in Figure~\ref{Fig:Figure6}, the intrinsic profiles of two models that reproduce Cloud-9's central column density. These are the same models presented in Figure~\ref{Fig:Figure5}: a ~\citetalias{Benitez-Llambay2017} RELHIC inhabiting a $\Lambda$CDM dark matter halo (green solid line) and an isothermal model (red dashed line), with sound speed matching the broadening of the VLA-D H\,{\sc i} 21-cm line. Both models reach the mean density of the universe at infinity. 

The reason why the isothermal model, as opposed to the~\citetalias{Benitez-Llambay2017} model, reproduces observations is readily apparent in the top right panel. The isothermal RELHIC matches Cloud-9's column density largely because its neutral hydrogen profile is intrinsically more extended than that of the~\citetalias{Benitez-Llambay2017} model. Although the central H\,{\sc i} density of the isothermal RELHIC is almost an order of magnitude lower than that of the~\citetalias{Benitez-Llambay2017} model, the larger physical extent of the neutral gas in the isothermal model compensates for this deficit, resulting in a similar central column density profile for both models. Furthermore, the more extended H\,{\sc i} distribution in the isothermal RELHIC leads to a more extended H\,{\sc i} column density profile, thus favoring a good match with observations.

The larger radial extent of the H\,{\sc i} profile compared to the~\citetalias{Benitez-Llambay2017} model arises from intrinsic differences between the two models in both temperature and density, as demonstrated in the left panels of Figure~\ref{Fig:Figure6}. Notably, compared to the adopted isothermal model,~\citetalias{Benitez-Llambay2017} RELHICs are colder in the center and hotter in the outskirts, favoring a steeper density gradient and a sharper decline of the H\,{\sc i} fraction with radius (see bottom right panel). Also, although the isothermal model has a lower neutral hydrogen fraction in the center compared to the~\citetalias{Benitez-Llambay2017} model, this deficit is compensated by the larger inner hydrogen density the system develops in comparison. 

Finally, we note that Figure~\ref{Fig:Figure6} explicitly shows that the~\citetalias{Benitez-Llambay2017} RELHICs matching Cloud-9's inner column density profile are colder than observed. It is thus not entirely surprising that these models fail to reproduce Cloud-9's column density profile in detail if the broadening of the H\,{\sc i} line traces the gas temperature of the neutral phase.

Despite our exercise highlighting the sensitivity of the assumed equation of state on the RELHICs' H\,{\sc i} density profile, it is remarkable to see that a simple model, with parameters motivated by observations, can reproduce Cloud-9's column density profile. 

Hence, we conclude that Cloud-9's detailed column density profile is consistent with an isothermal $\Lambda$CDM RELHIC whose temperature matches observations. However, we stress that taken at face value, the system appears to be in tension with the RELHIC model introduced by \citetalias{Benitez-Llambay2017}. This mismatch may simply reflect the fact that Cloud-9 is not truly in isolation, localized departures from the adopted UVB, or departures from the equation of state for the gas.

Until further optical observations confirm the existence of a luminous counterpart associated with the system, our VLA-D observations do not exclude the possibility that Cloud-9 is a RELHIC.

\subsection{Cloud-9 as a Local Group Dwarf Galaxy Analog}
\label{Sec:Leo T}

We now turn our attention to the hypothesis that Cloud-9 is a Local Group dwarf galaxy counterpart at the M94 distance, a case that rests on solid grounds for at least two reasons.

Firstly, results from hydrodynamical simulations and analytic considerations indicate that the majority of systems with dark matter mass similar to that inferred for Cloud-9 host luminous galaxies in their center. Indeed, ~\citetalias{Benitez-Llambay2017}~\citepalias[see also][and references therein]{Sawala2016, Benitez-Llambay2020} finds that more than $90 \%$ of the halos with present-day masses, $M_{200} > 5 \times 10^{9} M_{\odot}$, contain stars, making it likely that Cloud-9 is a luminous galaxy rather than a RELHIC. This idea is reinforced by the fact that quiescent low–mass galaxies today in cosmological hydrodynamical simulations display gaseous halos and total~H\,{\sc i} masses similar to those inferred for Cloud-9 and RELHICs~\citep[e.g.,][]{Pereira-Wilson2023, Herzog2023}.

Secondly, current photometric limits do not rule out the presence of a faint stellar counterpart at Cloud-9's location. For example, the surface brightness limits of the DESI Legacy Survey (LS) reported by~\citet[][]{Martinez-Delgado2023}, namely $29.15$, $28.73$, and $27.52$ $\rm mag \ arcsec^{-2}$ in the $g, r$, and $z$ bands, respectively, place large upper limits to the stellar mass of any luminous component associated with Cloud-9, as discussed by~\citetalias{Zhou2023}. In particular, these authors estimate that Cloud-9's luminous counterpart would not be more massive than $\approx 10^{5} \ M_{\odot}$. This maximum mass is well above the minimum stellar mass of some faint dwarfs observed in the Local Group~\citep[see, e.g.,][for a recent review on the faintest nearby galaxies]{Simon2019}, which makes us wonder whether there is any known nearby faint dwarf galaxy with properties similar to those inferred for Cloud-9 that might be missed in current observations of this field.

\subsubsection{Cloud-9 as a Leo T Lookalike}

Interestingly, Leo T  –a local dwarf galaxy lying at the outskirts of the Milky Way at a distance $d_{\rm LeoT} \approx 420 \rm \ kpc$– shares striking similarities with the properties derived for Cloud-9. The stellar mass of Leo T is $M_{\rm gal} \approx 10^{5} \ M_{\odot}$~\citep{Irwin2007}. The galaxy displays old and young stellar populations but no signs of ongoing star formation or H\,{\sc ii} regions at present despite having a total H\,{\sc i} mass, $M_{\rm HI} = 4.1 \times 10^{5} \ M_{\odot}$~\citep[][and references therein]{Irwin2007, Weisz2012, Adams2018}. Moreover, as we explore next, it would be difficult to detect its stellar counterpart at the fiducial distance of Cloud-9 in the DESI footprint.

To determine whether it is sensible to assume that Cloud-9 has a luminous counterpart similar to or fainter than Leo T, we consider a random realization of Leo T placed at Cloud-9's fiducial distance. To construct the photometric model, we use Leo T's structural parameters and the photometry for 182 individual stars reported by~\cite{Irwin2007}, who find that stars with colors, $(g-r) < 1$, and magnitudes, $m_{g} < 25$, are distributed in projection according to the following Plummer profile:\footnote{These parameters correspond to those listed in Table 1 of ~\cite{Irwin2007}, and reproduce the Plummer profile depicted in Figure 5 of that paper. However, we note that the central density value quoted by~\cite{Irwin2007} is erroneously expressed in magnitudes instead of $\rm arcsec^{-2}$.} 
\begin{equation}
    N(R) = 26.9 {\rm \ arcsec^{-2}} \left [ 1 + \left ( \displaystyle\frac{R}{1.4 \rm \ arcmin}\right)^2 \right ], 
\end{equation}

\begin{figure}
\includegraphics[width=\columnwidth]{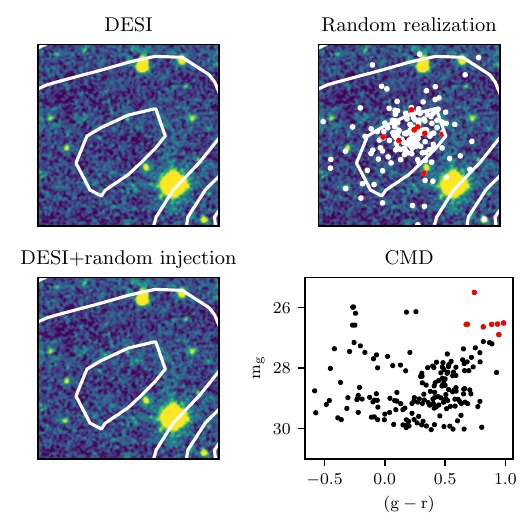}
        \caption{The top left panel shows the two innermost Cloud-9's H\,{\sc i} column density isocontours superimposed on a DESI LS cutout in the $g$ band. The top right panel shows the same plus a random realization of Leo T's best Plummer fit at Cloud-9's fiducial distance and centered at the peak column density (white dots). The bottom left panel shows this Plummer realization injected into the DESI LS data, in which we assigned magnitudes to the individual stars from the CMD shown in the bottom right panel and convolved with the DESI LS PSF of this field. For reference, the white open circles indicate the location of the eight brightest stars redder than $(g-r) = 0.5$ (red dots in the CMD panel).}
    \label{Fig:Figure7}
\end{figure}

One realization is shown in the top right panel of Figure~\ref{Fig:Figure6}, in which the white dots display the location of the stars for which we have photometric information, randomly sampled from the Plummer profile that best fits Leo T. The image in the background corresponds to a DESI LS cutout in the $g$ band. For reference, the panels are $2\times 2$ arcmin, and the solid lines indicate the innermost column density isocontours shown in Figure~\ref{Fig:Figure2}. We place Leo T's random realization at Cloud-9's distance and assign magnitudes to the stars from the color-magnitude diagram (CMD) displayed in the bottom right panel of the same figure. The CMD data was also taken from the work of~\cite{Irwin2007} and shifted from a distance of $420 \rm \ kpc$ to Cloud-9's distance, $d=4.66 \rm \ Mpc$.\footnote{Although~\cite{Irwin2007} report that Leo T's CMD corresponds to the stars in the inner $1.8$ arcmin from the center, we find that this assumption results in a wrong normalization for the reported Plummer profile. Therefore, we do not enforce a distance constraint to populate our Plummer realization, which results in a perfect match to~\cite{Irwin2007}'s results and a self-consistent normalization for the profile.} 

Detecting Leo T's tip of the red giant branch at Cloud-9's distance would thus require photometric observations deeper than $m_{g} \gtrsim 26$ magnitudes in the $g$ band, above which distinguishing stars from the expected large number of unresolved external galaxies becomes challenging with ground-based facilities~\cite[see, e.g.,][and references therein]{Soumagnac2015}.

To further highlight the difficulty of observing the stellar counterpart of a Leo T-type dwarf at Cloud-9's distance, we inject our random realization into the DESI LS cutout centered at Cloud-9's location and convolve it with the corresponding point-spread-function (PSF). The top left panel of Figure~\ref{Fig:Figure6} shows the original cutout in the g-band whereas the bottom left panel shows the same field after injecting Leo T's model. A quick visual inspection and comparison with the original cutout demonstrates that the brightest stars (we indicate the location of the eight brightest stars redder than $(g-r)>0.5$ with red symbols in the right-hand panels) are too faint to be seen in DESI.

This exercise demonstrates that detecting a Leo T-like galaxy at Cloud-9's location would be challenging with DESI imaging. Indeed, our exercise suggests that none of the Leo T stars can be detected with current data. Furthermore, the inability to see a potential Leo-T candidate in this cutout agrees with our measurement of the $5\sigma$ PSF magnitude detection limit at Cloud-9's location of approximately 25.6 in the g band.

Hence, we conclude that should Cloud-9 have an optical counterpart centered at its peak column density, its luminosity should be either comparable to or fainter than Leo T if the galaxy has a similar extent. This argument places a robust upper limit to the total stellar mass of any luminous counterpart associated with Cloud-9, $M_{\rm gal, CL9} \lesssim 10^5 \ M_{\odot}$, which aligns with previous estimates. 

How does Leo T's neutral hydrogen distribution compare with that of Cloud-9? To answer this question and carry out a meaningful comparison we mock-observe Leo T with the VLA-D ``clean'' beam, assuming it is at Cloud-9's distance. To this end, we first derive Leo T's intrinsic H\,{\sc i} column density profile by heuristically ``deconvolving'' the total averaged profile observed with the Westerbork Synthesis Radio Telescope (WSRT)~\citep{Adams2018}. 

The top panel of Figure~\ref{Fig:Figure7} shows that the following model reproduces~\cite{Adams2018} detailed observations after convolving it with the reported WSRT beam:\footnote{The WSRT beam for this observations is $(\theta_{M}, \theta_m)$ = $(57.3'', 15.7'')$~\citep[see Sec. 2 of][]{Adams2018}.} 
\begin{equation}
    N_{\rm HI} = 3.1 \times 10^{20} {\rm \ cm^{-2}} \exp \left [ -\left ( \displaystyle\frac{R}{0.21 \rm \ kpc } \right )^{3/2} \right ],
    \label{Eq:LeoTIntrinsic}
\end{equation}
We use this model to place Leo T's H\,{\sc i} component at Cloud-9's distance, after which we convolve it with the VLA-D ``clean'' beam. 

Comparing the mock-observed Leo T VLA-D column density profile (orange line in the bottom panel) with that of Cloud-9 (black symbols with shaded regions displaying $1\sigma$ uncertainties) reveals that both systems share a similar central column density value but Cloud-9 is more extended than Leo T, possibly due to Cloud-9's perturbed morphology.  

\begin{figure}
	\includegraphics[width=\columnwidth]{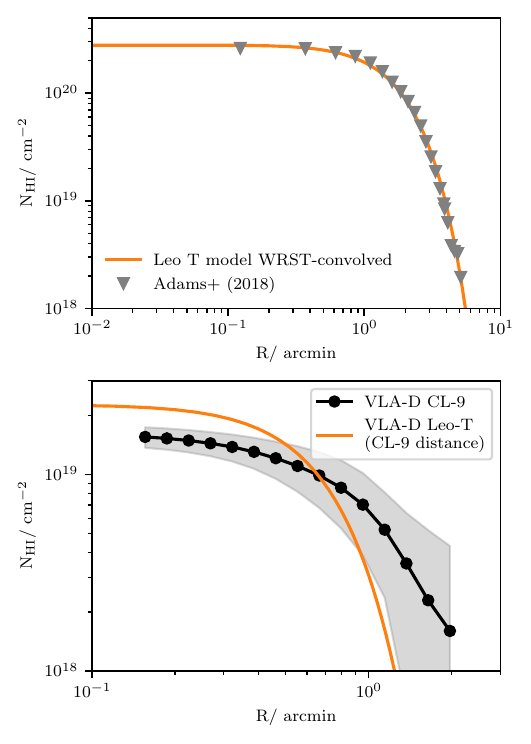}
        \caption{The top panel shows Leo T's total averaged H\,{\sc i} column density profile measured by~\cite{Adams2018} (grey triangles) together with our adopted model (Eq.~\ref{Eq:LeoTIntrinsic}) placed at Leo T's distance and convolved with the WRST beam (orange solid line). The bottom panel shows the observed Cloud-9's VLA-D H\,{\sc i} profile with $1-\sigma$ uncertainties (black circles and shaded region, respectively), together with our Leo T model placed at Cloud-9's distance, $d = 4.66 \rm \ Mpc$, convolved with the VLA-D beam.}
    \label{Fig:Figure8}
\end{figure}

The broad similarities between the gaseous components of Leo T and Cloud-9, combined with the limits of current optical surveys at Cloud-9's location, make it difficult to rule out the presence of a luminous galaxy counterpart with luminosity comparable to or fainter than Leo T. Moreover, the high expected frequency of luminous galaxies in halos with masses similar to that derived for Cloud-9 suggests that Cloud-9 is more likely an analog of Leo T rather than a starless RELHIC. Deep photometric observations of this field with the Hubble Space Telescope could help resolve individual stars fainter than 26 magnitudes, revealing an associated faint galaxy or tightening current constraints on the stellar mass of any luminous counterpart associated with Cloud-9. 
%__________________________________________________________________

\section{Discussion}
\label{Sec:Discussion}

While Cloud-9 appears round and symmetric in FAST observations, our new VLA-D data reveal that the system has a more complex morphology than previously anticipated. As discussed in Section~\ref{Sec:Results}, Cloud-9's neutral hydrogen reservoir appears perturbed compared to a spherical distribution centered at the peak column density, showing compression towards the west and expansion towards the east. Additionally, taken at face value, Cloud-9's detailed H\,{\sc i} column density profile is more extended than expected for the~\citetalias{Benitez-Llambay2017} RELHICs matching its central column density value, and also more extended than Leo T, a local dwarf galaxy with similar inferred properties. However, these features do not immediately rule out the RELHIC or Leo T-like counterpart interpretation for the system.

Firstly, if Cloud-9 were a RELHIC, its perturbed morphology could simply be ascribed to ram pressure caused by its motion relative to a gaseous background —a well-known process affecting galaxies that fall onto massive companions~\cite[e.g.][and references therein]{Gunn1972, Abadi1999}, and low-mass systems in isolation as they interact with the cosmic web~\citep{Benitez-Llambay2013, Herzog2023}. Moreover, the H\,{\sc i} content of some relatively isolated dwarf galaxies within our Local Group also displays signs of ram pressure~\citep[see, e.g.,][]{McConnachie2007}. 

The relative proximity of Cloud-9 to the massive galaxy M94 ($52$ arcmin in projection, corresponding to a projected distance of about $\approx 70$ kpc) makes this interpretation plausible. 

Given these issues, it is then not entirely surprising that the~\citetalias{Benitez-Llambay2017} model –designed to describe the gas distribution of RELHICs in an otherwise unperturbed background– fails to describe Cloud-9's gas distribution. Whether or not ram pressure can give rise to a~\citetalias{Benitez-Llambay2017} RELHIC resembling Cloud-9 is, however, uncertain. There are currently no theoretical predictions regarding the detailed distribution of the neutral hydrogen content of RELHICs undergoing ram pressure stripping.

Ram pressure-stripped starless halos have been identified in simulations by~\citetalias{Benitez-Llambay2017}, who noted that some of these systems, even when not necessarily satellites of a massive companion, lose their gas through cosmic web stripping. This process arises from the interaction between the small gas reservoir of RELHICs and the gas in the filaments and sheets of the cosmic web. These authors also observed that RELHICs efficiently lose their loosely bound gas when encountering even mild gas overdensities in the field. This results in a dichotomy where the gas content of starless halos is either in hydrostatic equilibrium with their host dark matter halo or is lost completely. However, this dichotomy between RELHICs and ``gas-free'' halos may stem from the limited resolution of these simulations\footnote{~\citetalias{Benitez-Llambay2017}'s simulations cannot reliably track the gas content of systems with masses, $M_{\rm gas} << 10^5 \ M_{\odot}$.}, making it unclear whether or not the central H\,{\sc i} structure of ram pressure-stripped RELHICs, as modeled by~\citetalias{Benitez-Llambay2017}, would resemble that of Cloud-9. Further high-resolution numerical simulations, tailored to resolve the gas-stripping process in RELHICs, may shed light on these uncertainties.

On the other hand, the intrinsic properties of the~\citetalias{Benitez-Llambay2017} RELHICs shown in Figure~\ref{Fig:Figure6} demonstrate that these systems are colder in the central regions compared to the inferred temperature for Cloud-9. Given the sensitivity of the models to the adopted temperature, as highlighted in Figure~\ref{Fig:Figure6}, it is not entirely surprising that the~\citetalias{Benitez-Llambay2017} model fails to match observations. Within the RELHIC interpretation, the different temperature for Cloud-9 compared to the expectations from the~\citetalias{Benitez-Llambay2017} model may originate from either variations in the local intensity of the UVB, or perturbations induced by ram pressure, or both. Moreover, FAST and GBT observations suggest the presence of diffuse emission missed in our VLA-D observations, displaying a larger broadening. These facts, together with Cloud-9's perturbed morphology, may be signaling the presence of a mechanism responsible for increasing the temperature of the system compared to what is naively expected. 

Modeling Cloud-9 as an isothermal RELHIC, with a temperature matching our VLD-D observations, has proven successful. Specifically, we have demonstrated that Cloud-9's column density profile can be accurately modeled by assuming an underlying dark matter distribution consistent with that expected for $\Lambda$CDM halos. This success supports the idea that Cloud-9 is a starless dark matter halo in near hydrostatic equilibrium with its dark matter component. The success of this model in matching Cloud-9's column density profile stems from the hotter inner temperature assumed compared to the~\citetalias{Benitez-Llambay2017} RELHIC model. Understanding in detail the mechanisms that could increase the inner temperature of RELHICs and make them depart from the~\citetalias{Benitez-Llambay2017} equation of state requires, however, more advanced modeling.

Further progress and a deeper understanding of this system in the context of RELHICs may be achieved through very high-resolution simulations of individual RELHICs inhabiting $\Lambda$CDM halos, with sufficient resolution to trace their inner H\,{\sc i} reservoir as it becomes stripped by external ambient gas moving relative to the system, modeling also the effect of the UVB self-consistently using radiative transfer calculations to estimate the ionization state of the gas as it expands. Understanding the telltale signatures of ram-pressure stripping in RELHICs, particularly regarding the central distribution of the neutral hydrogen content and resulting temperature structure, will undoubtedly provide robust theoretical foundations to better interpret our observations. 

Additionally, understanding Cloud-9’s outer morphology with deeper radio observations may prove useful in assessing its external structure, whereas high-resolution radio observations will help constrain its inner morphology and assess whether the system is in equilibrium.

If Cloud-9 were a luminous galaxy similar to Leo T, but with its gas being stripped by ram pressure, our analysis suggests that deeper observations might detect its stars. The best chances of identifying a galaxy associated with Cloud-9 would be through deep HST observations that can resolve faint individual stars and distinguish them from background sources. The galaxy could be detected and characterized by reaching $5\sigma$ PSF magnitudes, $m_{g} \approx 30$. For instance, observations reaching $m_{g} \lesssim 29$ would detect Leo T's 20 brightest stars at Cloud-9's distance and reveal the tip of the red giant branch. Observations reaching $m_{g} \lesssim 30$ would identify approximately 30 stars.

Finally, detecting and resolving the stars of a galaxy at Cloud-9's location would be remarkable, as this would become the faintest galaxy known at that distance. In turn, characterizing its stellar populations would help place robust constraints on the distance of the system and benchmark current estimates based on the redshift of the H\,{\sc i} 21-cm line.

%__________________________________________________________________
\section{Conclusions}
\label{Sec:Conclusions}

We have presented VLA-D interferometric observations of a recently discovered H\,{\sc i} cloud (Cloud-9) located on the outskirts of M94 and lacking a luminous counterpart. Our analysis follows up on the discovery of the system by~\citetalias{Zhou2023} and the subsequent analysis by~\citetalias{Benitez-Llambay2023}, who showed that the system is consistent with being a $\Lambda$CDM starless dark matter halo filled with gas in hydrostatic equilibrium and in thermal equilibrium with the external UVB. The main results of our work may be summarized as follows: 

\begin{itemize}
    \item{Although FAST observations by \citetalias{Zhou2023} reveal Cloud-9 as a largely spherical system, our new interferometric VLA-D observations indicate a more complex morphology. We find that the system appears perturbed, exhibiting features consistent with compression of its H\,{\sc i} content towards the west and expansion towards the east. We have discussed the possibility that these features originate from ram pressure between the inner gas content of the system and ambient gas associated with M94.}
    \item{Unlike~\citetalias{Karunakaran2024}, we find that Cloud-9's 21-cm H\,{\sc i} line appears symmetric and Gaussian, with a very narrow broadening, $W_{50}=12 \pm 1 \rm \ km \ s^{-1}$. This value sharply contrasts with that derived by~\citetalias{Zhou2023} and~\citetalias{Karunakaran2024} using the FAST and GBT instruments, respectively. We speculate that the differences may arise from our interferometric observations missing some diffuse emission component. However, our observations compellingly demonstrate the presence of a prominent, dynamically cold, neutral hydrogen component.}
    \item{Cloud-9 does not exhibit signs of ordered rotation within the sampled scales. The lack of rotation suggests that if the system is in equilibrium, it is the gas pressure that maintains this balance. Assuming equilibrium, the extent of Cloud-9's gas distribution and the narrow H\,{\sc i} 21-cm line profile indicates the presence of a large amount of gravitational mass that is not neutral hydrogen, suggesting the system is either a dark matter-dominated galaxy or a starless RELHIC.}
    \item{Taken at face value, Cloud-9's H\,{\sc i} column density profile is more extended than the~\citetalias{Benitez-Llambay2017}$+\Lambda$CDM RELHIC models. As noted in earlier work by~\citetalias{Benitez-Llambay2023} and confirmed here, the system appears much more extended in projection than RELHICs. If the system is indeed in equilibrium with a dark matter halo and thermal equilibrium with the~\cite{Haardt2012} UVB, this would imply halo properties distinct from those predicted by $\Lambda$CDM. However, this discrepancy may also stem from Cloud-9's outer perturbed morphology or discrepancies in the temperature of the system compared to the modeling. Assessing these scenarios requires further theoretical work to understand the impact of ram-pressure stripping on the gaseous structure of RELHICs, particularly on the centrally concentrated H\,{\sc i} component and its temperature.}
    \item{Despite the difficulties of the \citetalias{Benitez-Llambay2017} model in describing Cloud-9's column density profile, we find that a model consisting of an isothermal gaseous sphere with a sound speed of $\approx 12 \ \rm km \ s^{-1}$ matching the broadening of the 21-cm H\,{\sc i} line successfully reproduces Cloud-9's column density profile. This is achievable if the cloud is in hydrostatic equilibrium with $\Lambda$CDM halos of mass $3.9 \lesssim M_{200} / 10^{9} \ M_{\odot} \lesssim 4.3$. This model is arguably better suited for Cloud-9 given observational constraints and the uncertain intensity of the UVB at Cloud-9's location.}
    \item{The high expected frequency of luminous galaxies in halos as massive as inferred for Cloud-9 ($M_{200} > 3 \times 10^{9} \ M_{\odot}$) suggests that Cloud-9 could be a luminous galaxy. Moreover, we find that Cloud-9's gas properties resemble those of Leo T, a nearby H\,{\sc i}-rich dwarf galaxy in the Local Group that is not currently forming stars. Specifically, their similar inner H\,{\sc i} column density suggests that Cloud-9 may be a Leo T-like galaxy. This possibility is consistent with current imaging DESI LS data in which it is challenging to detect a Leo T analog. Our analysis places a realistic upper limit to the stellar mass of a luminous counterpart associated with Cloud-9 of about $M_{\rm gal, CL9} \lesssim 10^5 \ M_{\odot}$.}
\end{itemize}

To make progress in the characterization of Cloud-9, we recommend further observations in both optical and radio wavelengths. Given the high likelihood of Cloud-9 hosting a luminous galaxy, we suggest deep photometric observations with either the Hubble Space Telescope or the James Web Space Telescope, targeting PSF magnitudes $m_{g} > 30$. The ability of these observations to sample the tip of the red giant branch of a Leo T analog at Cloud-9's distance would allow us to detect the galaxy or significantly lower the limits on the stellar mass of such a counterpart if no stars are detected. Given the importance of RELHICs as a test for the $\Lambda$CDM model, we believe these efforts are well justified. Additionally, our analysis indicates that any galaxy associated with Cloud-9 will likely become the faintest known to date at that distance, and certainly, not the first in the local volume discovered through follow-up observations of 21-cm emission~\cite[see, e.g.,][for the discovery of the faint Leo P galaxy]{Giovanelli2013}. 

Secondly, to better understand the gaseous structure of the system, we recommend efforts on two fronts. First, further radio observations should aim to resolve both larger and smaller scales, potentially helping constrain the temperature profile of the system. Observing the system with VLA in the B or C-array configuration would help decrease the beam's impact and better resolve the central regions, crucial for assessing Cloud-9's internal equilibrium. Second, additional observations with FAST would allow us to study the outer diffuse emission further and understand the true extent of Cloud-9, helping the assessment of departures from equilibrium or signs of coherent rotation on large scales. 

Finally, further theoretical work tailored to understand the gas-stripping process in RELHICs, particularly on the central H\,{\sc i} reservoir and its temperature, may prove useful and help better interpret Cloud-9's morphology. Current simulations that resolve the gas content of RELHICs lack sufficient resolution to characterize the structure of affected systems in detail. Until these analyses are available, it will be difficult to determine whether the tension between Cloud-9 and $\Lambda$CDM RELHICs, as modeled by~\citetalias{Benitez-Llambay2017}, challenges $\Lambda$CDM or simply reflects Cloud-9's perturbed morphology.

In conclusion, Cloud-9 still proves to be a fascinating system that, regardless of its nature, likely probes the faint edge of galaxy formation. We strongly advocate for deep optical and radio observations, alongside high-resolution simulations that can resolve the H\,{\sc i} stripping process of RELHICs and faint dwarfs. These efforts will not only clarify the nature of this system but also provide valuable insights into the understanding of the clustering of dark matter and the process of galaxy formation at the smallest scales. 

\begin{acknowledgements}
The National Radio Astronomy Observatory is a facility of the National Science Foundation operated under cooperative agreement by Associated Universities, Inc.
A.B.L. acknowledges support from the European Research Council (ERC) under the European Union's Horizon 2020 research and innovation program (GA 101026328). This study is supported by the Italian Ministry for Research and University (MUR) under Grant ``Progetto Dipartimenti di Eccellenza 2023-2027'' (BiCoQ). 
\end{acknowledgements}

\bibliography{bibliography}

\begin{thebibliography}{}
\expandafter\ifx\csname natexlab\endcsname\relax\def\natexlab#1{#1}\fi
\providecommand{\url}[1]{\href{#1}{#1}}
\providecommand{\dodoi}[1]{doi:~\href{http://doi.org/#1}{\nolinkurl{#1}}}
\providecommand{\doeprint}[1]{\href{http://ascl.net/#1}{\nolinkurl{http://ascl.net/#1}}}
\providecommand{\doarXiv}[1]{\href{https://arxiv.org/abs/#1}{\nolinkurl{https://arxiv.org/abs/#1}}}

\bibitem[{{Abadi} {et~al.}(1999){Abadi}, {Moore}, \& {Bower}}]{Abadi1999}
{Abadi}, M.~G., {Moore}, B., \& {Bower}, R.~G. 1999, \mnras, 308, 947,
  \dodoi{10.1046/j.1365-8711.1999.02715.x}

\bibitem[{{Adams} {et~al.}(2013){Adams}, {Giovanelli}, \& {Haynes}}]{Adams2013}
{Adams}, E. A.~K., {Giovanelli}, R., \& {Haynes}, M.~P. 2013, \apj, 768, 77,
  \dodoi{10.1088/0004-637X/768/1/77}

\bibitem[{{Adams} \& {Oosterloo}(2018)}]{Adams2018}
{Adams}, E. A.~K., \& {Oosterloo}, T.~A. 2018, \aap, 612, A26,
  \dodoi{10.1051/0004-6361/201732017}

\bibitem[{{Benitez-Llambay} \& {Frenk}(2020)}]{Benitez-Llambay2020}
{Benitez-Llambay}, A., \& {Frenk}, C. 2020, \mnras, 498, 4887,
  \dodoi{10.1093/mnras/staa2698}

\bibitem[{{Benitez-Llambay} \& {Navarro}(2023)}]{Benitez-Llambay2023}
{Benitez-Llambay}, A., \& {Navarro}, J.~F. 2023, \apj, 956, 1,
  \dodoi{10.3847/1538-4357/acf767}

\bibitem[{{Ben{\'\i}tez-Llambay} {et~al.}(2013){Ben{\'\i}tez-Llambay},
  {Navarro}, {Abadi}, {Gottl{\"o}ber}, {Yepes}, {Hoffman}, \&
  {Steinmetz}}]{Benitez-Llambay2013}
{Ben{\'\i}tez-Llambay}, A., {Navarro}, J.~F., {Abadi}, M.~G., {et~al.} 2013,
  \apjl, 763, L41, \dodoi{10.1088/2041-8205/763/2/L41}

\bibitem[{{Ben{\'\i}tez-Llambay} {et~al.}(2017){Ben{\'\i}tez-Llambay},
  {Navarro}, {Frenk}, {Sawala}, {Oman}, {Fattahi}, {Schaller}, {Schaye},
  {Crain}, \& {Theuns}}]{Benitez-Llambay2017}
{Ben{\'\i}tez-Llambay}, A., {Navarro}, J.~F., {Frenk}, C.~S., {et~al.} 2017,
  \mnras, 465, 3913, \dodoi{10.1093/mnras/stw2982}

\bibitem[{{Benson} {et~al.}(2002){Benson}, {Frenk}, {Lacey}, {Baugh}, \&
  {Cole}}]{Benson2002}
{Benson}, A.~J., {Frenk}, C.~S., {Lacey}, C.~G., {Baugh}, C.~M., \& {Cole}, S.
  2002, \mnras, 333, 177, \dodoi{10.1046/j.1365-8711.2002.05388.x}

\bibitem[{{Blumenthal} {et~al.}(1984){Blumenthal}, {Faber}, {Primack}, \&
  {Rees}}]{Blumenthal1984}
{Blumenthal}, G.~R., {Faber}, S.~M., {Primack}, J.~R., \& {Rees}, M.~J. 1984,
  \nat, 311, 517, \dodoi{10.1038/311517a0}

\bibitem[{{Bullock} {et~al.}(2000){Bullock}, {Kravtsov}, \&
  {Weinberg}}]{Bullock2000}
{Bullock}, J.~S., {Kravtsov}, A.~V., \& {Weinberg}, D.~H. 2000, \apj, 539, 517,
  \dodoi{10.1086/309279}

\bibitem[{{Dickey}(1990)}]{Dickey1990}
{Dickey}, J.~M. 1990, in Astrophysics and Space Science Library, Vol. 161, The
  Interstellar Medium in Galaxies, ed. J.~{Thronson}, Harley~A. \& J.~M.
  {Shull}, 473--481, \dodoi{10.1007/978-94-009-0595-5_19}

\bibitem[{{Giovanelli} {et~al.}(2013){Giovanelli}, {Haynes}, {Adams}, {Cannon},
  {Rhode}, {Salzer}, {Skillman}, {Bernstein-Cooper}, \&
  {McQuinn}}]{Giovanelli2013}
{Giovanelli}, R., {Haynes}, M.~P., {Adams}, E. A.~K., {et~al.} 2013, \aj, 146,
  15, \dodoi{10.1088/0004-6256/146/1/15}

\bibitem[{{Gunn} \& {Gott}(1972)}]{Gunn1972}
{Gunn}, J.~E., \& {Gott}, J.~Richard, I. 1972, \apj, 176, 1,
  \dodoi{10.1086/151605}

\bibitem[{{Haardt} \& {Madau}(2012)}]{Haardt2012}
{Haardt}, F., \& {Madau}, P. 2012, \apj, 746, 125,
  \dodoi{10.1088/0004-637X/746/2/125}

\bibitem[{{Herzog} {et~al.}(2023){Herzog}, {Ben{\'\i}tez-Llambay}, \&
  {Fumagalli}}]{Herzog2023}
{Herzog}, G., {Ben{\'\i}tez-Llambay}, A., \& {Fumagalli}, M. 2023, \mnras, 518,
  6305, \dodoi{10.1093/mnras/stac3282}

\bibitem[{{Hoeft} {et~al.}(2006){Hoeft}, {Yepes}, {Gottl{\"o}ber}, \&
  {Springel}}]{Hoeft2006}
{Hoeft}, M., {Yepes}, G., {Gottl{\"o}ber}, S., \& {Springel}, V. 2006, \mnras,
  371, 401, \dodoi{10.1111/j.1365-2966.2006.10678.x}

\bibitem[{{Irwin} {et~al.}(2007){Irwin}, {Belokurov}, {Evans}, {Ryan-Weber},
  {de Jong}, {Koposov}, {Zucker}, {Hodgkin}, {Gilmore}, {Prema}, {Hebb},
  {Begum}, {Fellhauer}, {Hewett}, {Kennicutt}, {Wilkinson}, {Bramich},
  {Vidrih}, {Rix}, {Beers}, {Barentine}, {Brewington}, {Harvanek},
  {Krzesinski}, {Long}, {Nitta}, \& {Snedden}}]{Irwin2007}
{Irwin}, M.~J., {Belokurov}, V., {Evans}, N.~W., {et~al.} 2007, \apjl, 656,
  L13, \dodoi{10.1086/512183}

\bibitem[{{Karunakaran} \& {Spekkens}(2024)}]{Karunakaran2024}
{Karunakaran}, A., \& {Spekkens}, K. 2024, Research Notes of the American
  Astronomical Society, 8, 24, \dodoi{10.3847/2515-5172/ad1ee6}

\bibitem[{{Ludlow} {et~al.}(2016){Ludlow}, {Bose}, {Angulo}, {Wang},
  {Hellwing}, {Navarro}, {Cole}, \& {Frenk}}]{Ludlow2016}
{Ludlow}, A.~D., {Bose}, S., {Angulo}, R.~E., {et~al.} 2016, \mnras, 460, 1214,
  \dodoi{10.1093/mnras/stw1046}

\bibitem[{{Mart{\'\i}nez-Delgado} {et~al.}(2023){Mart{\'\i}nez-Delgado},
  {Roca-F{\`a}brega}, {Mir{\'o}-Carretero}, {G{\'o}mez-Flechoso}, {Rom{\`a}n},
  {Donatiello}, {Schmidt}, {Lang}, {Akhlaghi}, \&
  {Hanson}}]{Martinez-Delgado2023}
{Mart{\'\i}nez-Delgado}, D., {Roca-F{\`a}brega}, S., {Mir{\'o}-Carretero}, J.,
  {et~al.} 2023, \aap, 669, A103, \dodoi{10.1051/0004-6361/202244832}

\bibitem[{{McClure-Griffiths} {et~al.}(2023){McClure-Griffiths},
  {Stanimirovi{\'c}}, \& {Rybarczyk}}]{McClure-Griffiths2023}
{McClure-Griffiths}, N.~M., {Stanimirovi{\'c}}, S., \& {Rybarczyk}, D.~R. 2023,
  \araa, 61, 19, \dodoi{10.1146/annurev-astro-052920-104851}

\bibitem[{{McConnachie} {et~al.}(2007){McConnachie}, {Venn}, {Irwin}, {Young},
  \& {Geehan}}]{McConnachie2007}
{McConnachie}, A.~W., {Venn}, K.~A., {Irwin}, M.~J., {Young}, L.~M., \&
  {Geehan}, J.~J. 2007, \apjl, 671, L33, \dodoi{10.1086/524887}

\bibitem[{{McMullin} {et~al.}(2007){McMullin}, {Waters}, {Schiebel}, {Young},
  \& {Golap}}]{McMullin2007}
{McMullin}, J.~P., {Waters}, B., {Schiebel}, D., {Young}, W., \& {Golap}, K.
  2007, in Astronomical Society of the Pacific Conference Series, Vol. 376,
  Astronomical Data Analysis Software and Systems XVI, ed. R.~A. {Shaw},
  F.~{Hill}, \& D.~J. {Bell}, 127

\bibitem[{{Navarro} {et~al.}(1996){Navarro}, {Frenk}, \& {White}}]{Navarro1996}
{Navarro}, J.~F., {Frenk}, C.~S., \& {White}, S. D.~M. 1996, \apj, 462, 563,
  \dodoi{10.1086/177173}

\bibitem[{{Nebrin} {et~al.}(2023){Nebrin}, {Giri}, \& {Mellema}}]{Nebrin2023}
{Nebrin}, O., {Giri}, S.~K., \& {Mellema}, G. 2023, \mnras, 524, 2290,
  \dodoi{10.1093/mnras/stad1852}

\bibitem[{{Okamoto} \& {Frenk}(2009)}]{Okamoto2009}
{Okamoto}, T., \& {Frenk}, C.~S. 2009, \mnras, 399, L174,
  \dodoi{10.1111/j.1745-3933.2009.00748.x}

\bibitem[{{Okamoto} {et~al.}(2008){Okamoto}, {Gao}, \& {Theuns}}]{Okamoto2008}
{Okamoto}, T., {Gao}, L., \& {Theuns}, T. 2008, \mnras, 390, 920,
  \dodoi{10.1111/j.1365-2966.2008.13830.x}

\bibitem[{{Pereira-Wilson} {et~al.}(2023){Pereira-Wilson}, {Navarro},
  {Ben{\'\i}tez-Llambay}, \& {Santos-Santos}}]{Pereira-Wilson2023}
{Pereira-Wilson}, M., {Navarro}, J.~F., {Ben{\'\i}tez-Llambay}, A., \&
  {Santos-Santos}, I. 2023, \mnras, 519, 1425, \dodoi{10.1093/mnras/stac3633}

\bibitem[{{Rahmati} {et~al.}(2013){Rahmati}, {Pawlik}, {Rai{\v{c}}evi{\'c}}, \&
  {Schaye}}]{Rahmati2013}
{Rahmati}, A., {Pawlik}, A.~H., {Rai{\v{c}}evi{\'c}}, M., \& {Schaye}, J. 2013,
  \mnras, 430, 2427, \dodoi{10.1093/mnras/stt066}

\bibitem[{{Roberts}(1962)}]{Roberts1962}
{Roberts}, M.~S. 1962, \aj, 67, 437, \dodoi{10.1086/108752}

\bibitem[{{Sawala} {et~al.}(2016){Sawala}, {Frenk}, {Fattahi}, {Navarro},
  {Theuns}, {Bower}, {Crain}, {Furlong}, {Jenkins}, {Schaller}, \&
  {Schaye}}]{Sawala2016}
{Sawala}, T., {Frenk}, C.~S., {Fattahi}, A., {et~al.} 2016, \mnras, 456, 85,
  \dodoi{10.1093/mnras/stv2597}

\bibitem[{{Simon}(2019)}]{Simon2019}
{Simon}, J.~D. 2019, \araa, 57, 375,
  \dodoi{10.1146/annurev-astro-091918-104453}

\bibitem[{{Soumagnac} {et~al.}(2015){Soumagnac}, {Abdalla}, {Lahav}, {Kirk},
  {Sevilla}, {Bertin}, {Rowe}, {Annis}, {Busha}, {Da Costa}, {Frieman},
  {Gaztanaga}, {Jarvis}, {Lin}, {Percival}, {Santiago}, {Sabiu}, {Wechsler},
  {Wolz}, \& {Yanny}}]{Soumagnac2015}
{Soumagnac}, M.~T., {Abdalla}, F.~B., {Lahav}, O., {et~al.} 2015, \mnras, 450,
  666, \dodoi{10.1093/mnras/stu1410}

\bibitem[{{Sykes} {et~al.}(2019){Sykes}, {Fumagalli}, {Cooke}, {Theuns}, \&
  {Ben{\'\i}tez-Llambay}}]{Sykes2019}
{Sykes}, C., {Fumagalli}, M., {Cooke}, R., {Theuns}, T., \&
  {Ben{\'\i}tez-Llambay}, A. 2019, \mnras, 487, 609,
  \dodoi{10.1093/mnras/stz1234}

\bibitem[{{Wang} {et~al.}(2024){Wang}, {Lin}, {Yang}, {Staveley-Smith},
  {Walter}, {Wang}, {Wang}, {Battisti}, {Catinella}, {Chen}, {Cortese},
  {Fisher}, {Ho}, {Ji}, {Jiang}, {Kauffmann}, {Kong}, {Liu}, {Shao}, {Wang},
  {Wang}, \& {Wang}}]{Wang2024}
{Wang}, J., {Lin}, X., {Yang}, D., {et~al.} 2024, arXiv e-prints,
  arXiv:2404.09422, \dodoi{10.48550/arXiv.2404.09422}

\bibitem[{{Weisz} {et~al.}(2012){Weisz}, {Zucker}, {Dolphin}, {Martin}, {de
  Jong}, {Holtzman}, {Dalcanton}, {Gilbert}, {Williams}, {Bell}, {Belokurov},
  \& {Evans}}]{Weisz2012}
{Weisz}, D.~R., {Zucker}, D.~B., {Dolphin}, A.~E., {et~al.} 2012, \apj, 748,
  88, \dodoi{10.1088/0004-637X/748/2/88}

\bibitem[{{Wilson} {et~al.}(2009){Wilson}, {Rohlfs}, \&
  {H{\"u}ttemeister}}]{Wilson2009}
{Wilson}, T.~L., {Rohlfs}, K., \& {H{\"u}ttemeister}, S. 2009, {Tools of Radio
  Astronomy}, \dodoi{10.1007/978-3-540-85122-6}

\bibitem[{{Xu} {et~al.}(2023){Xu}, {Zhu}, {Yu}, {Zhang}, {Liu}, {Ai}, \&
  {Jiang}}]{Xu2023}
{Xu}, J.-L., {Zhu}, M., {Yu}, N., {et~al.} 2023, \apjl, 944, L40,
  \dodoi{10.3847/2041-8213/acb932}

\bibitem[{{Zhang} {et~al.}(2024){Zhang}, {Zhu}, {Jiang}, {Cheng}, {Wang},
  {Wang}, {Xu}, {Liu}, {Yu}, {Qian}, {Yu}, {Ai}, {Jing}, {Xu}, {Liu}, {Guan},
  {Sun}, {Yang}, {Huang}, {Hao}, \& {FAST Collaboration}}]{Zhang2024}
{Zhang}, C.-P., {Zhu}, M., {Jiang}, P., {et~al.} 2024, Science China Physics,
  Mechanics, and Astronomy, 67, 219511, \dodoi{10.1007/s11433-023-2219-7}

\bibitem[{{Zhou} {et~al.}(2023){Zhou}, {Zhu}, {Yang}, {Yu}, {Yuan}, {Jiang}, \&
  {Xi}}]{Zhou2023}
{Zhou}, R., {Zhu}, M., {Yang}, Y., {et~al.} 2023, \apj, 952, 130,
  \dodoi{10.3847/1538-4357/acdcf5}

\end{thebibliography}
\bibliographystyle{aasjournal}

\end{document}